\documentclass[aps,prl,twocolumn,showpacs,amsmath,amssymb,superscriptaddress,10pt,floatfix]{revtex4-1}

\usepackage{amsmath}
\usepackage{amsfonts}
\usepackage{amssymb}
\usepackage{color}
\usepackage{units}
\usepackage{verbatim}
\usepackage{graphicx}

\newcommand{\beq}{\begin{equation}}
\newcommand{\eeq}{\end{equation}}
\newcommand{\bqa}{\begin{eqnarray}}
\newcommand{\eqa}{\end{eqnarray}}

\newcommand{\bra}[1]{\left\langle{#1}\right|}
\newcommand{\ket}[1]{\left|{#1}\right\rangle}

\renewcommand{\section}[1]{\textit{#1}.---}

\begin{document}

\title{Experimentally Violating Bell Inequalities Without Complete Reference Frames}

\author{Matthew S. Palsson}
\affiliation{Centre for Quantum Computation and Communication Technology (Australian Research Council), Centre for Quantum Dynamics, Griffith University, Brisbane, 4111, Australia}

\author{Joel J. Wallman}
\affiliation{School of Physics, The University of Sydney, Sydney, New South Wales, 2006, Australia}

\author{Adam J. Bennet}
\affiliation{Centre for Quantum Computation and Communication Technology (Australian Research Council), Centre for Quantum Dynamics, Griffith University, Brisbane, 4111, Australia}

\author{G. J. Pryde}\email{G.Pryde@griffith.edu.au}
\affiliation{Centre for Quantum Computation and Communication Technology (Australian Research Council), Centre for Quantum Dynamics, Griffith University, Brisbane, 4111, Australia}

\begin{abstract}

We experimentally demonstrate, using qubits encoded in photon polarization, that if two parties share a single reference direction and use locally orthogonal measurements they will \textit{always} violate a Bell inequality, up to experimental deficiencies. This contrasts with the standard view of Bell inequalities in which the parties need to share a complete reference frame for their measurements. Furthermore, we experimentally demonstrate that as the reference direction degrades the probability of violating a Bell inequality decreases smoothly to 39.7$\pm$0.1\% in the limiting case that the observers do not share a reference direction. This result promises simplified distribution of entanglement between separated parties, with applications in fundamental investigations of quantum physics and tasks such as quantum communication. 
\end{abstract}

 \maketitle

\section{Introduction}
Shared entanglement between two or more parties is an important resource in the development of technologies and protocols exploiting the properties of quantum systems. The correlations between entangled systems may be harnessed to implement quantum information and communication protocols \cite{Nielsen2000} such as quantum teleportation \cite{bennett93,horo96} and quantum key distribution (QKD) \cite{ekert91,masanes09}. Of particular note is Ekert's QKD protocol~\cite{ekert91}, in which two parties can only be confident that no eavesdroppers have intercepted a shared key if they can violate a Bell inequality \cite{gisin02}. However, in order to maximally violate a Bell inequality, the two parties must share a complete reference frame, i.e., be able to perfectly align their measurements relative to each other. If they do not share a complete reference frame, infinite communication is required to establish one, which is experimentally infeasible \cite{bart03,rud03}.

There exist a number of challenges in employing finite (necessarily imperfect) reference frames, including their fidelity, the operational cost of establishing them, and their degradation if comprised of quantum systems \cite{BRSRMP07}. Information encoded in relative degrees of freedom can be used to avoid the need for a reference frame \cite{cabello}, but such schemes add significant complexity in terms of encoding, measurement, and noise suppression, so they are undesirable in many practical situations.

Here, we demonstrate experimentally that two parties can always violate a Bell inequality if they share a \textit{single} reference direction, up to experimental imperfections. We also investigate how the probability of violating a Bell inequality decreases as the shared reference direction degrades. This degradation of a shared reference frame between two parties is of particular interest for long range quantum communication protocols whose security is guaranteed by the violation of a Bell inequality. 

In our experiment, we consider the propagation of entangled photons as a natural means of remotely sharing entanglement. In such a scenario, it is natural to consider situations where two parties, Alice and Bob, share a single reference direction but not a complete reference frame. An example is where they share a single reference polarization basis (such as horizontal/vertical)---set by some shared anisotropy such as gravity or the known axis of a polarization maintaining fiber---as they communicate over a channel between two stations. A second example is that they may share knowledge of the direction of propagation of the light through a non-birefringent medium, but no knowledge of the relative orientation of their apparatus about that axis, as in the case of line-of-sight communication between satellites which may be moving or rotating.  While the unknown polarization rotation can be determined and can be approximately compensated in a specific experimental run, this is difficult to accomplish over long distances. If this compensating step can be eliminated, then it makes, e.g., implementing Ekert's QKD protocol substantially easier \cite{laing}.

As we use a polarization encoding of qubits, calibrating the relative alignments of measurements at a single site is straightforward. We use half-wave plates (HWP) to switch between measurement bases at each site and so need only consider the case where the two measurements at each site are mutually unbiased, i.e., correspond to perpendicular directions on the Poincar\'{e} sphere. Each party chooses their two measurements to be in the plane of the Poincar\'{e} sphere whose normal vector corresponds, up to a sign, to their approximation to the shared reference direction. If the parties share the direction perfectly, then their approximation is exact and so their measurements are coplanar in the Poincar\'{e} sphere. It has recently been shown that, theoretically, these measurements will always violate a Bell inequality, \cite{joel}. Here we investigate this protocol experimentally and violate a Bell inequality $99.3\pm 0.3\%$ of the time. However, if they do not share a common direction (i.e.\ their local approximation to a shared direction is a vector distributed uniformly over the surface of a sphere), then they will violate a Bell inequality 41.3\% of the time in the absence of experimental deficiencies \cite{liang10}. In our experiment, $39.7\pm 0.1\%$ of cases lead to a violation.

\section{Theoretical Background}
In our experimental investigation, we use the CHSH-Bell inequality \cite{CHSH}. The CHSH inequality imposes an upper bound on the correlation function between Alice and Bob's measurement outcomes when they measure all four combinations of two pairs of observables on each system---one combination per trial---over many trials on identically prepared bipartite systems in any locally causal model. Specifically,

\beq\label{eq:CHSH_parameter}
\lvert\langle S_{CHSH}\rangle\rvert = \lvert\langle XP + ZP + XQ - ZQ\rangle\rvert\leq2 
\eeq

\noindent in any locally causal theory, where $X, Z \in\{\pm1\}$ are the result of a pair of measurements made by Alice, and similarly for $P, Q \in\{\pm1\}$ for Bob. However, measurements on the maximally entangled two qubit spin singlet state $\ket{\Psi^{-}}= (\ket{0}_{A}\otimes\ket{1}_{B} - \ket{1}_{A}\otimes\ket{0}_{B})/\sqrt{2}$ may violate this inequality. When Alice chooses to measure from the maximally complementary Pauli bases $\hat{X}\equiv\ket{0}\bra{1}+\ket{1}\bra{0}$ and $\hat{Z}\equiv\ket{0}\bra{0}-\ket{1}\bra{1}$, and Bob from the maximally complementary bases $\hat{P}\equiv-(\hat{Z}+\hat{X})/\sqrt{2}$ and $\hat{Q}\equiv(\hat{Z}-\hat{X})/\sqrt{2}$, then the Bell-CHSH parameter $|\langle\hat{S}_{CHSH}\rangle|=2\sqrt{2}$, providing a violation of the inequality by a factor of $\sqrt{2}$. This is the maximal violation of the CHSH inequality allowed by quantum mechanics \cite{cirel80}. 

Note, however, that if Alice and Bob used the above measurements except Bob relabeled his measurement outcomes, i.e., measured in the bases $\hat{P}'\equiv-\hat{P}$ and $\hat{Q}'\equiv-\hat{Q}$ so that the $\pm1$ eigenstates are swapped, then the Bell-CHSH parameter $|\langle\hat{S}_{CHSH}\rangle|=0$ is consistent with a locally causal theory. Consequently, to identify a violation of Eq.~\eqref{eq:CHSH_parameter}, we need to consider all such  relabelings of measurement outcomes. However, if Alice and Bob employ the same pair of maximally complementary settings (e.g.\ $\hat{X},\hat{Z}$) then the CHSH inequality is satisfied for any relabeling of measurement outcomes.  Consequently, Alice and Bob's measurements need to be aligned in particular ways in order for them to violate the CHSH inequality.

In this paper we will investigate the probability of violating a CHSH inequality when the the parties do not share a complete reference frame and so are unable to perform the measurements required to obtain the maximal violation of the CHSH inequality. To visualize the choices of measurements, we will use the mapping from observables $\hat{O}$ to unit vectors $\vec{r}(\hat{O})\in\mathbb{R}^3$ given by
\begin{align}
\hat{O} = \vec{r}(\hat{O})\cdot\vec{\sigma}	\,,
\end{align}
where $\vec{\sigma} = (X,Y,Z)$ is the vector of Pauli matrices. Alice and Bob's observables are chosen in their local coordinate systems, i.e., on two Poincar\'{e} spheres that are related by some rotation. Alice and Bob share a reference frame if they know the relation between their respective Poincar\'{e} spheres. We assume that Alice chooses the measurements $\hat{Z}$ and $\hat{X}$ on her Poincar\'{e} sphere, while Bob chooses the measurements $\hat{P}$ and $\hat{Q}$ on his. In this work, the Pauli directions map to polarisation states as follows: horizontal (H) and vertical (V) polarizations are the $\pm Z$ eigenstates; diagonal (D) and antidiagonal (A) are the $\pm X$ eigenstates; and right and left hand circular (R and L, respectively) are the $\pm Y$ eigenstates. We denote polarizations representing the $\hat{P}$ and $\hat{Q}$ eigenstates as $P^{\pm}$ and $Q^{\pm}$ respectively.

As the singlet state has no angular momentum in the Schwinger representation (i.e., is the two qubit $j=0$ state), it is invariant under joint rotations of Alice and Bob's Poincar\'{e} spheres. Therefore we can hold Bob's Poincar\'{e} sphere to be fixed and assume that any evolution only induces a rotation of Alice's Poincar\'{e} sphere.

To describe the situation where the two observers share a single reference direction, we consider (without loss of generality) the example where they share exact knowledge of direction of propagation in a non-birefringent medium. Thus, they share knowledge of the plane of linear polarizations---and therefore its normal, the $\pm Y$ Bloch direction---although they have no knowledge of their relative orientation in the $XZ$ plane. In terms of the state transformation, we represent this by a rotation $R_Y(\theta)$ of an angle  $\theta$ around the $Y$ axis of the Bloch sphere \cite{note1}.

\begin{figure}[t]
 \centering
 \includegraphics[scale=0.33]{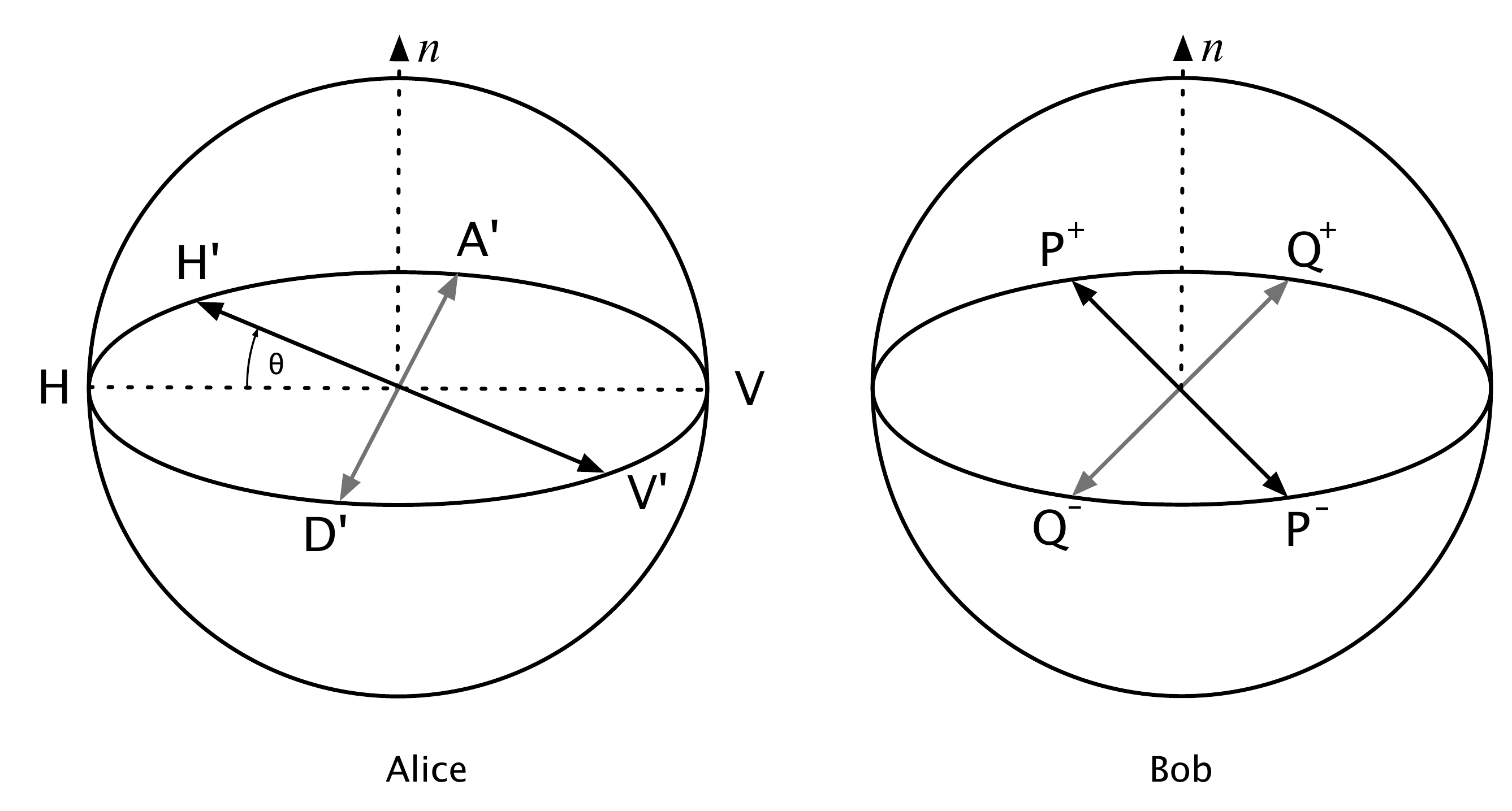}
 \caption{Alice and Bob's Poincar\'{e} spheres when they perfectly share the $Y$ direction. Alice's measurement bases are rotated from the canonical $H$ and $V$ polarization measurements (relative to Bob's Poincar\'{e} sphere) by an angle $\theta$ to new measurement directions denoted by primed (${}^\prime$) text.}
 \label{fig:CHSH}
\end{figure}

\begin{figure}[t]
 \centering
 \includegraphics[scale=0.33]{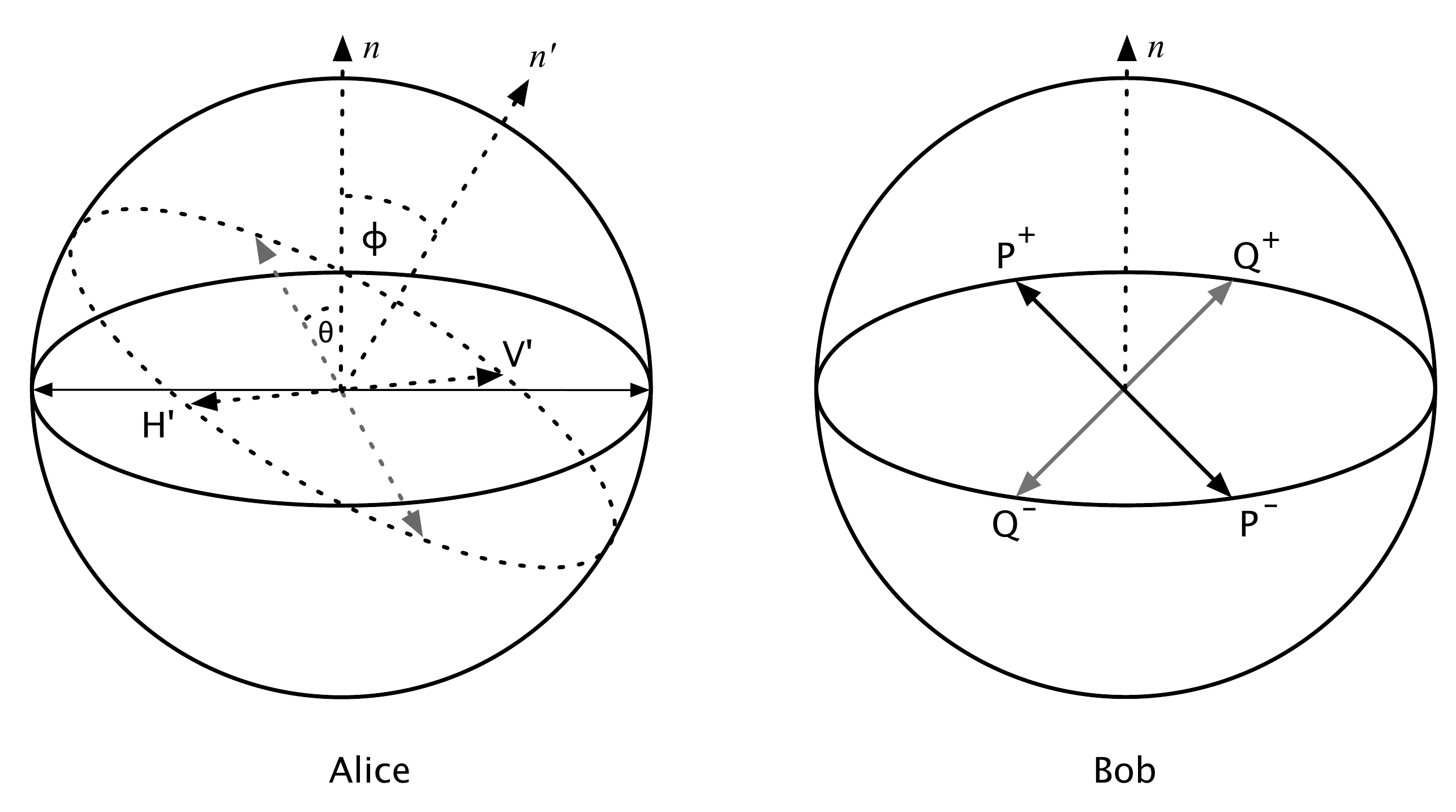}
 \caption{Alice and Bob's Poincar\'{e} spheres in the case where no reference direction is perfectly shared. Alice's new measurement plane is defined by rotations about three separate axes. Sampling across all three rotations allows for the nonlocal correlations shared between Alice and Bob to be investigated as the single shared reference direction degrades by angle $\phi$.}
 \label{fig:phitheta}
\end{figure}

However, it may be the case that Alice and Bob do not \textit{exactly} share the transmission direction, or that the R and L polarization states are otherwise not perfectly invariant under transmission. In this case, the overall evolution of Alice's Bloch sphere can be written as $R_y(\chi)R_z(\phi)R_y(\theta)$. This can be visualized as follows: first, rotate Alice's $H$ and $V$ measurements around the $Y$ axis by an angle $\theta$, as in Fig.~\ref{fig:CHSH}. Then rotate the $Y$ axis to the vector $n'=(-\sin\phi\cos\chi,\cos\phi,\sin\phi\sin\chi)$ as shown in Fig.~\ref{fig:phitheta}. However, it can be shown that the correlations between Alice and Bob's measurements only depend on the difference $\theta - \chi$, therefore we set $\chi=0$.

\section{Experimental Implementation}
The experimental setup is as follows. We generate polarization-entangled photon pairs by spontaneous parametric down conversion (SPDC) in a type-II periodically poled Potassium titanyl phosphate.  The crystal is phase matched at $\sim$12$^{\circ}$C to produce degenerate photon pairs at 820~nm from a 410~nm continuous-wave $<$1 mW diode laser pump. Measurements made using automated polarization analysis optics and single photon counting modules in each output arm of the source allows the two qubit density matrix to be reconstructed via quantum state tomography \cite{alt04}. The same measurement apparatus is used for the CHSH-style measurements in this experiment. Typical operation of this source produces $\sim$1500s$^{-1}$ states having fidelity 0.994$\pm$0.001 with the closest maximally entangled state~\cite{state}. The maximum value of the CHSH parameter we obtain is  2.81$\pm$0.01, only slightly below the maximum value of $2\sqrt{2}$ allowed by quantum mechanics \cite{cirel80} and consistent with the CHSH value of $|\langle S_{CHSH}\rangle|\approx 2.81$ we expect to achieve, given the measured state fidelity.

\begin{figure}[ht]
 \centering
 \includegraphics[scale=0.24]{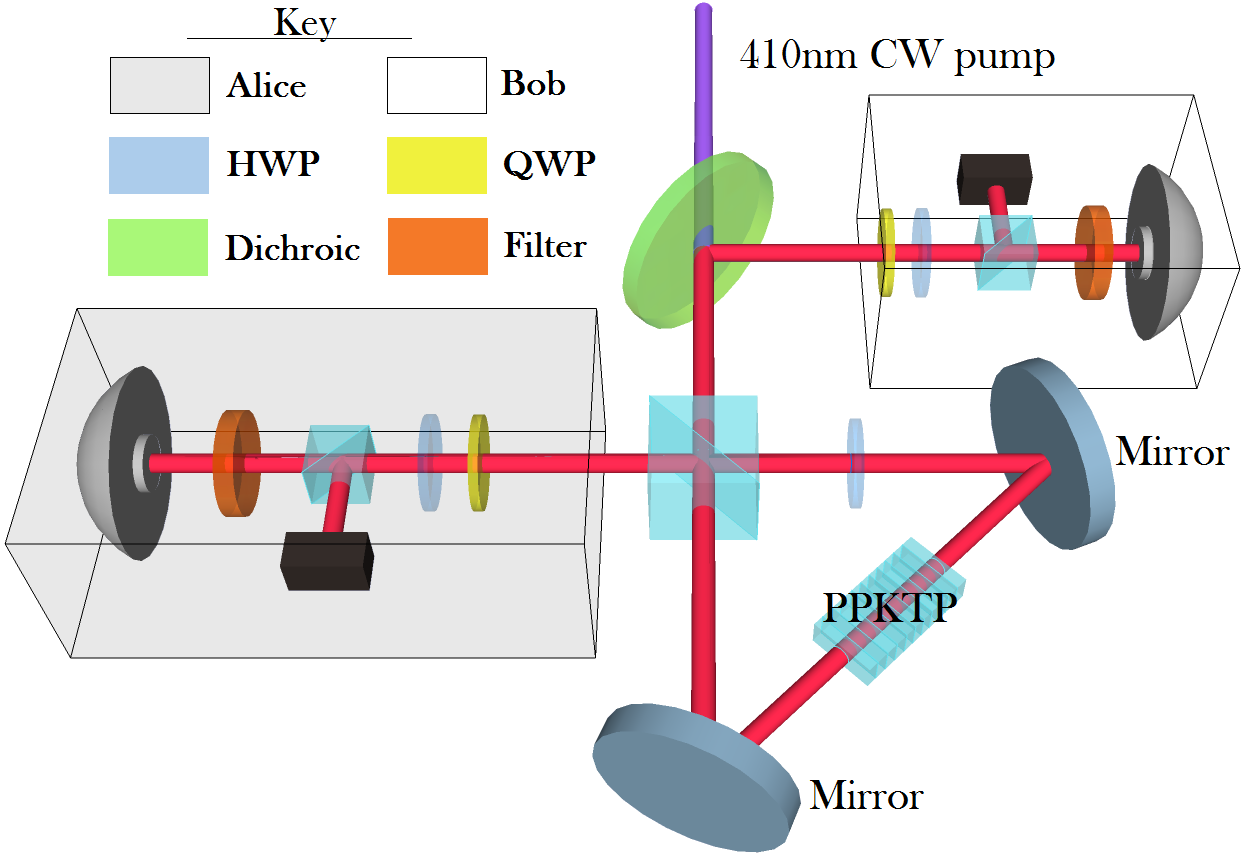}
 \caption{Experimental layout of the PPKTP SPDC Sagnac source and measurement apparatus. The PPKTP crystal is pumped bi-directionally in superposition, resulting in the maximally entangled $\ket{\Psi^{-}}$ state being generated at the output ports of a polarising beam splitter. Quarter and half wave plates are used in conjunction with Glan-Taylor prisms to measure the polarisation state of down converted photons. Interference and coloured glass filters are used to block background light.}
 \label{fig:exp}
\end{figure}

A key feature of this experiment is that we systematically sample from the full range of possible relative measurement directions (i.e., pairs $(\theta,\phi$) rather than performing a random sample. This is done by fixing $\phi=[0\,:\,10\,:\,90]$ \cite{1} and then calculating the CHSH parameter for each value of $\theta=[0\,:\,10\,:\,180]$. Note that this method of sampling does not give a uniform sampling over the surface of the sphere, which would assign probability $p(\phi)\propto \cos\phi$ to picking a point with polar angle $\phi$. To estimate the probability of violating the CHSH inequality for $\phi\in[0, 10t]$ using pairs of points $(\theta,\phi)$ with $\theta=[0\,:\,10\,:\,180]$ and $\phi=[0\,:\,10\,:\,10t]$, we evaluate
\begin{align}\label{eq:lower_bound}
p(t) = \sum_{s=0}^{t} \mu(s) f(10 s)
\end{align}
for an appropriate probability measure $\mu(s)$, where $f(\phi)$ is the fraction of the values of $\theta$ such that $(\theta,\phi)$ lead to a violation of the CHSH inequality. When $t=0$, i.e., Alice and Bob share the $Y$ direction perfectly, we set $\mu(0) = 1$. For $t> 0$, we set
\begin{align}
\mu(s) = \begin{cases} 0 & \text{if } s = 0	\,, \\
C[\cos(10s-10) - \cos(10s)] & \text{otherwise}\,, \end{cases}
\end{align}
where $C$ is a normalization factor.

When Alice and Bob share a direction perfectly, the theoretical prediction is that they will always violate a Bell inequality, except when their measurements are perfectly aligned (a set of zero measure in the space of relative settings), in which case the CHSH inequality is saturated. For states that are not quite maximally entangled, the CHSH inequality is satisfied for small regions around the completely aligned condition \cite{joel}. In our experiment, Alice and Bob perfectly share a direction when $\phi = 0$ and their measurements would be perfectly aligned when $\theta = 45^{\circ}$ or $135^{\circ}$ (in the absence of experimental imperfections). 

\begin{figure}[ht]
 \centering
 \includegraphics[scale=0.32]{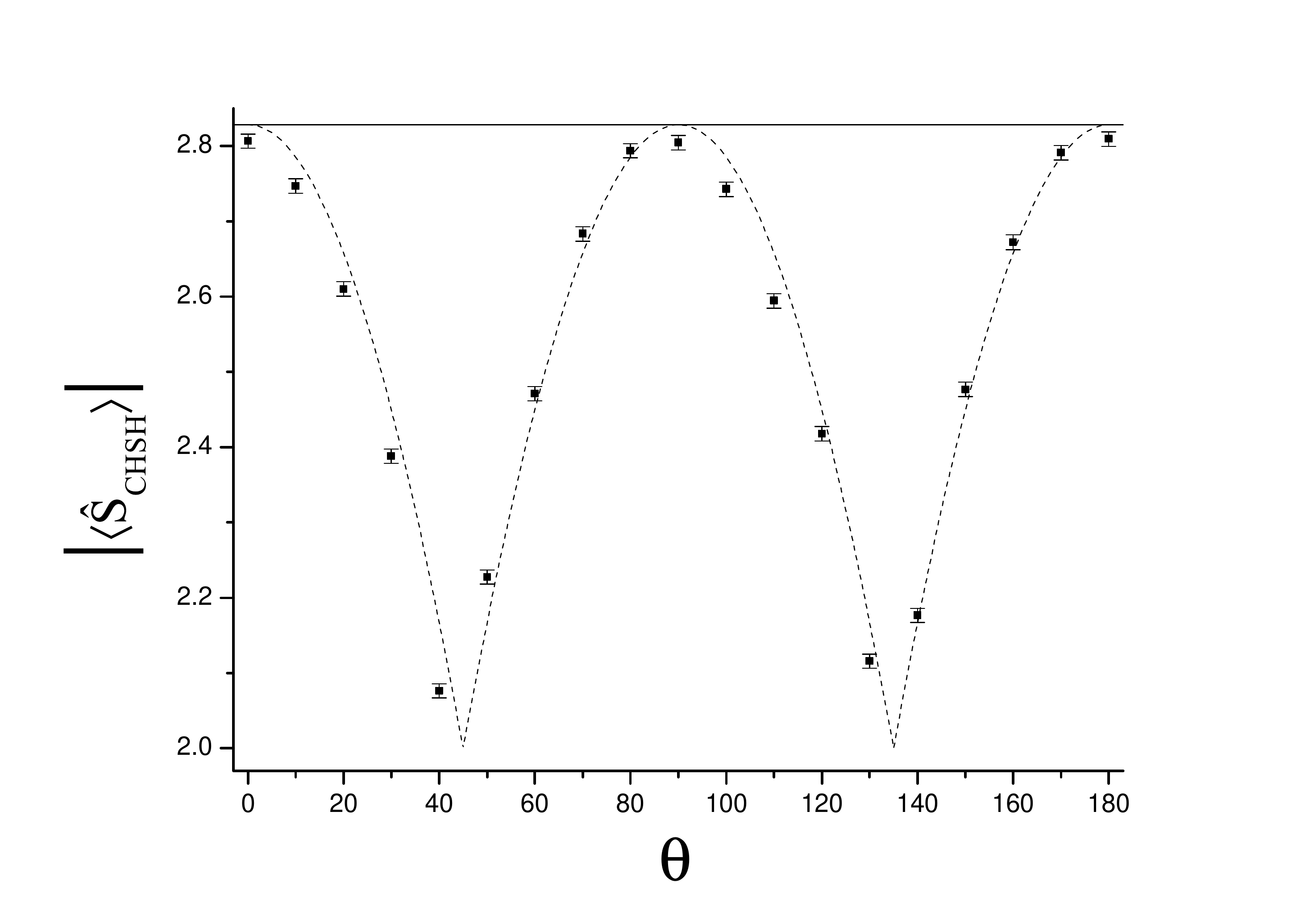}
 \caption{Experimentally measured outcomes of the CHSH parameter sampled in 10 degree steps of $\theta$, when Alice and Bob's $Y$ axes are completely aligned. The dashed line illustrates the theoretically calculated value for the prepared state.  The solid black line represents the maximum value of the CHSH parameter allowed by quantum mechanics.  Error bars are one standard deviation, derived from Poissonian counting statistics.}
 \label{fig:phi0}
\end{figure} 

\begin{figure}[t]
 \centering
 \includegraphics[scale=0.32]{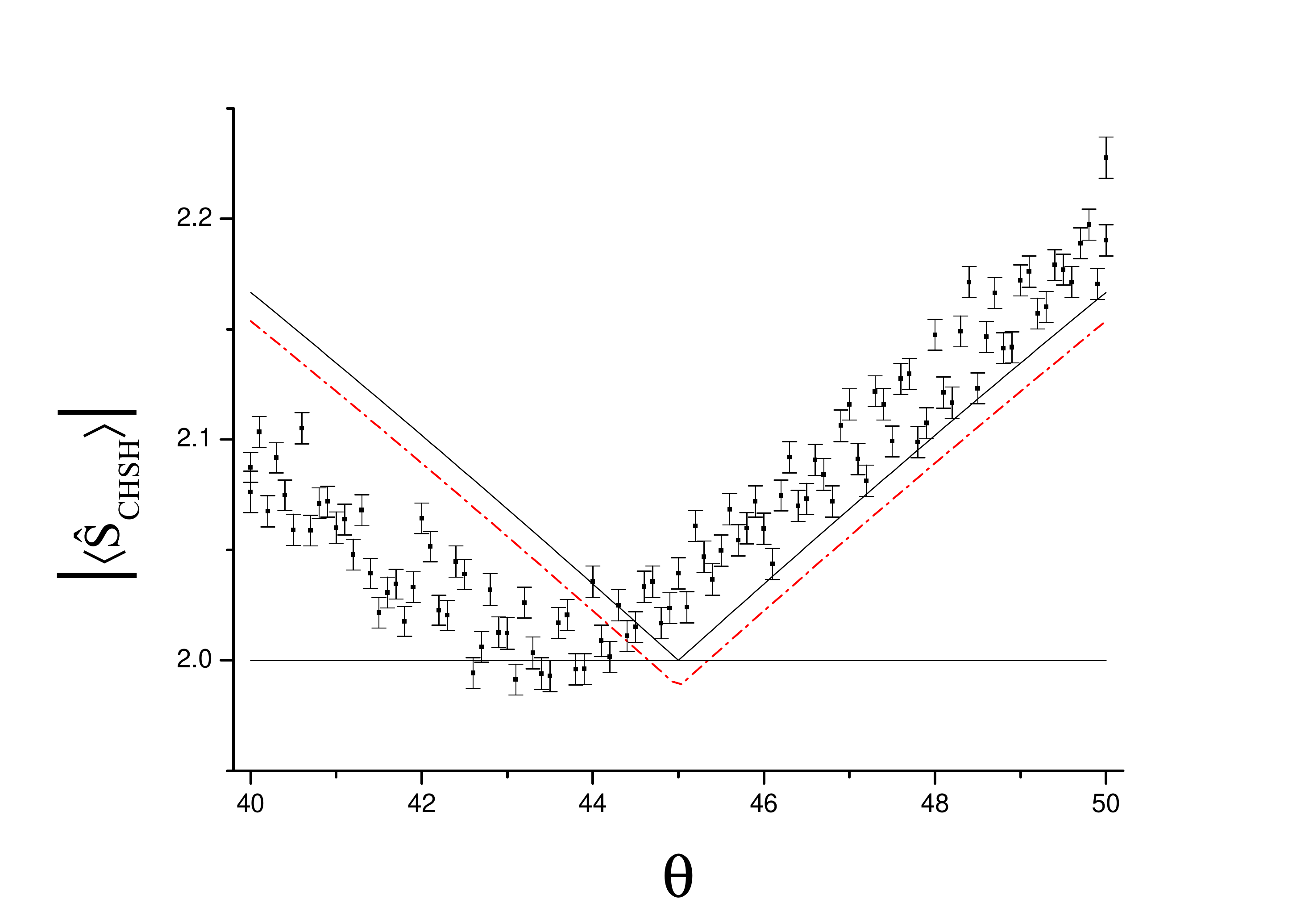}
 \caption{High resolution scan about the expected range of angles where the CHSH parameter will be at a minimum. Expected theoretical values for a perfect singlet state are shown by the solid black line, the dashed red line denotes theoretical values for a state with fidelity 0.994. All data points violate the CHSH inequality (2) to within uncertainty, except one data point. The expected minimum is shifted away, in angle, from the theoretically expected minimum due to wave plate and state imperfections.}
 \label{fig:blah}
\end{figure}

As shown in Figure \ref{fig:phi0}, violations were observed when $\phi = 0$ for all relative angles $\theta \in [0^\circ\,:\,10\,:\,180^\circ]$, with minimum values occuring near $\theta = 45^{\circ}$ and $\theta = 135^{\circ}$.  To examine these regions further, we collected the high angular resolution data shown in Figure \ref{fig:blah} (with steps of 0.1${}^\circ$ in $\theta$ in a small region centered around $\theta = 45^{\circ}$). We find that every data point violates the CHSH inequality by more than one standard deviation (as determined by photon counting statistics) except for some data points in a small region around the minima (i.e., $\theta\in[42.6^{\circ}\,:\,0.1^{\circ}\,:\,44.2^{\circ}]$). The theoretical minimum expectation value of the CHSH parameter for the given state is $\approx 1.988$, in agreement with the minimum measured CHSH parameter of $1.991\pm0.007$. The experimental minimum is shifted from the theoretical expected minima to $\approx 43^{\circ}$ due to slight imperfections in state preparation (which lead to unitary single-qubit rotations away from the ideal state), and small imperfections in wave plate retardance and settings. 

From our experimental data, we infer that in regions away from $\theta \approx 45^\circ$ and $\theta \approx 135^\circ$ that the CHSH inequality is always violated when $\phi = 0^{\circ}$. Assuming that the measurement statistics around $\theta \approx 135^\circ$ are the same as those around $\theta \approx 45^\circ$ (justified by the  theoretical symmetry which is also reflected in our measurement data), we infer that the CHSH inequality is violated with probability $99.3 \pm 0.3\%$. This agrees with the theoretically predicted value of $99.2\pm 0.1\%$ for states with the same degree of entanglement as the one we produce.

\begin{figure}[t]
 \centering
 \includegraphics[scale=0.32]{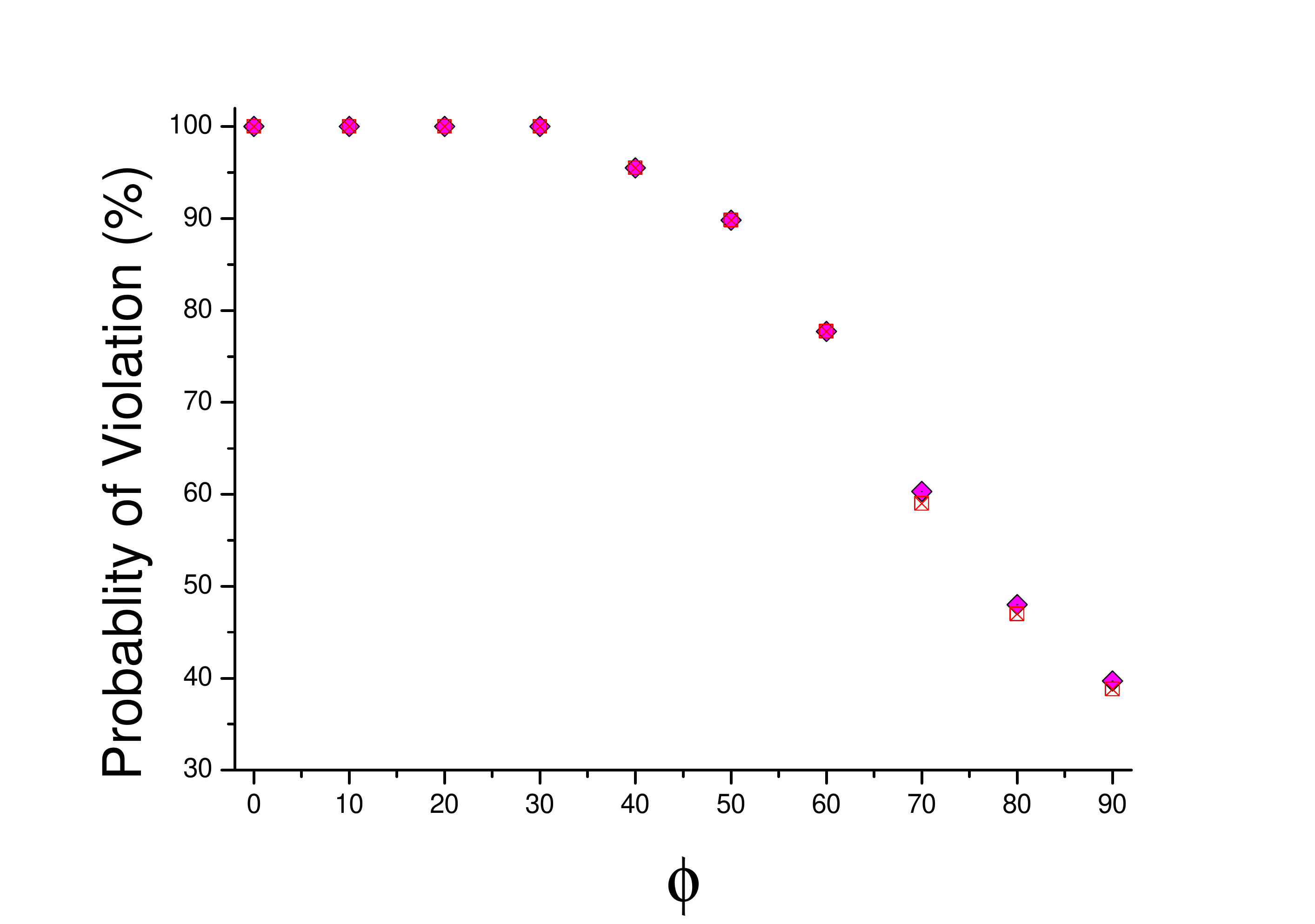}
 \caption{Experimentally measured probability of violating the CHSH inequality as alignment of Alice and Bob's $Y$ axis degrades by angle $\phi$ and the parameter $\chi$ is fixed at 0, while $\theta$ is sampled evenly. Purple diamonds represent that the mean values violate the CHSH inequality. Red crosses indicate the probability the mean is more than a standard deviation away from the bound.}
 \label{fig:chi0}
\end{figure}

As $\phi$ increases from 0, the extent to which Alice and Bob share a reference frame decreases. Our experimental results show that this leads to a corresponding decrease in the probability of violating a Bell inequality, as shown in Fig. \ref{fig:chi0}. When $\phi = 90^{\circ}$, Alice and Bob do not share a reference frame at all, so the scenario is equivalent to Alice and Bob independently choosing two maximally complementary bases each.  Using Eq. \eqref{eq:lower_bound}, we obtain an experimental lower bound of 39.7$\pm$0.1\%. This is in close agreement with the theoretical value of $40.3\pm 0.2 \%$ for states with the same degree of entanglement as the one we produce \cite{note2}.

\section{Conclusions}
We have experimentally demonstrated that a complete shared reference frame is not required for two remote parties to violate a Bell inequality. If the parties share just one measurement direction perfectly, they can almost always violate a Bell inequality perfectly with a maximally entangled state by each choosing two maximally complimentary measurements in the plane orthogonal to the shared direction in the Poincar\'{e} sphere.  Furthermore, even moderate errors in the alignment of the shared direction do not significantly reduce the probability of achieving a violation. Our work has direct application to quantum information protocols, such as quantum key distribution, that use single site measurements of entangled quantum states.

During the preparation of this manuscript, related theoretical and experimental work has shown that by making more complex measurements---increasing the number of measurements that each party chooses---can allow the observers to always violate a Bell inequality without any alignment of the reference frames \cite{shadbolt,joel2}.


\begin{thebibliography}{99}
\bibitem{note1} Equivalently, we could think of a measurement rotation by $-\theta$.
\bibitem{state} For our data, the closest maximally entangled state is related to the singlet $\ket{\Psi^-}$ by small local rotations.
\bibitem{note2} This probability was estimated numerically by determining the fraction of a sample of $10^7$ measurement directions that led to a violation of a Bell inequality and is consistent with the analytic value for the ideal case \cite{liang10}
\end{thebibliography}
\end{document}